\documentclass[
 reprint,
 amsmath,amssymb,
 aps,
prl
]{revtex4-1}

\usepackage{layout}

\usepackage{graphicx}
\usepackage{dcolumn}
\usepackage{bm}

\begin{document}



\title{Shot noise suppression in InGaAs/InGaAsP quantum channels}

\author{Yoshitaka Nishihara$^1$, Shuji Nakamura$^1$, Kensuke Kobayashi$^1$}
 \email{kensuke@scl.kyoto-u.ac.jp.}
\author{Teruo Ono$^1$, Makoto Kohda$^{2,3}$} 
\author{Junsaku Nitta$^2$}%
\affiliation{%
 $^1$Institute for Chemical Research, Kyoto University, Uji, Kyoto 611-0011, Japan\\
 $^2$Department of Materials Science, Tohoku University, 6-6-02 Aramaki-Aza Aoba, Aoba-ku, Sendai 980-8579,Japan\\
 $^3$PRESTO, Japan Science and Technology Agency, 4-1-8 Honcho, Kawaguchi, Saitama 332-0012, Japan
}%




\date{\today}

\begin{abstract}
We have measured the shot noise in a quantum point contact (QPC) fabricated by using InGaAs/
InGaAsP heterostructure, whose conductance can be electrically tuned by the gate voltages. The
reduced shot noise is observed when the QPC conductance equals to $N(2e^ 2/h)$ $(N=4, 5, \mathrm{and} \; 6)$,
which is the direct experimental evidence of the coherent quantized channel formation in the QPC.
The deviation of the observed Fano factor from the theory is explained by the electron heating effect
generated at the QPC.
\end{abstract}

\maketitle

InGaAs-based devices are attracting great interest these days since the InGaAs two-dimensional heterostructure possesses a strong spin-orbit interaction (SOI), which provides us ample opportunities to perform electrical spin generation \cite{MurakamiScience2003,KatoScience2004}, manipulation \cite{DattaAPL1989,KogaPRB2006}, and detection \cite{OhePRB2005} in solid state devices.
Among various InGaAs-based systems including selfassembled quantum dots and nano whiskers \cite{Nadj-PergeNature2010}, the twodimensional electron gas (2DEG) system is an ideal stage to address spin-dependent coherent quantum transport \cite{KogaPRB2006,NittaPRL1997,KogaPRL2002}.
Indeed, in 1980 s and 1990 s, mesoscopic systems represented by Aharonov-Bohm (AB) rings and the quantum point contacts (QPCs) \cite{WeesPRL1988,WharamJPC1988} on GaAs-based 2DEG have served to establish quantum transport \cite{DattaCUP1995}.
In the same way but in focusing more on the role of spins, the AB rings on InGaAs 2DEG
were studied to demonstrate the electron spin interference a few years ago \cite{BergstenPRL2006}.
Spin resolved quantized conductance in the InAs QPC was reported very recently \cite{DebrayNatNano2009}.
Thus, the quantum spin transport in the mesoscopic systems made of InGaAs 2DEG is now being started to be exploered \cite{KohdaPRB2010,EngelsPRB1997}, although there remains much to be addressed when compared to what have been done for those in GaAs-based 2DEG.

Shot noise is a powerful tool to study quantum transport in mesoscopic systems \cite{BlanterPR2000}.
For example, the shot noise measurement was performed for the QPC on GaAs 2DEG already in 1990 (Ref. \cite{DekkerPRL1991}) followed by the works by Reznikov {\it et al}. \cite{ReznikovPRL1995} and Kumar {\it et al}. \cite{KumarPRL1996}, who showed that the Fano factor is suppressed at the conductance plateau.
This shot noise suppression, which is essentially originated from the Pauli principle of electrons, is a significant consequence of the formation of the coherent quantized channels.
Moreover, the shot noise is expected to be useful to investigate coherent spin transport and spin correlation \cite{FevePRB2002,EguesPRL2002}.
However, no report on the shot noise has been available for the InGaAs-based QPCs \cite{MartinPRB2008,SchapersAPL2007,SimmondsAPL2008}.

Here, we report the shot noise study of the QPC made by using InGaAs/InGaAsP 2DEG, where strong SOI exists.\cite{KohdaPRB2010}
We found that the shot noise is reduced at the conductance plateaus of $N(2e^2/h)$ $(N=4, 5, \mathrm{and} \;6)$. 
This is the direct experimental evidence of the coherent quantized channels in the InGaAs-based QPC. 
We also discuss the heating effect to explain the deviation of the observed Fano factor from the conventional shot noise theory.
The present shot noise study is a step forward to further explore quantum spin transport in InGaAs-based systems.

\begin{figure*}
\centering
\includegraphics{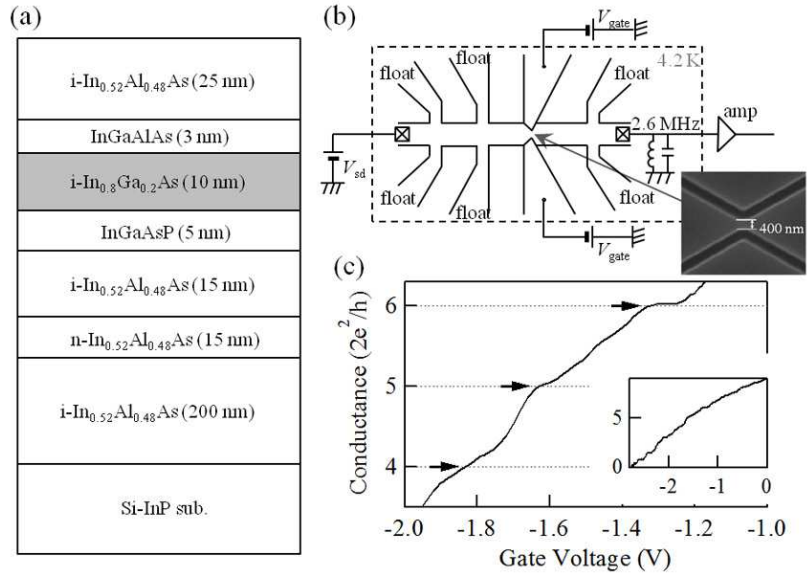}
\caption{(a) Schematic layer structure of an InGaAs/InGaAsP.
The 2DEG is formed in $\textrm{i-In}_{0.8} \textrm{Ga}_{0.2} \textrm{As}$ layer shown as gray color.
(b) Schematic diagram of the conductance and noise measurement setup.
The inset shows a scanning electron microscope image of the sample fabricated on 2DEG.
(c) QPC conductance at $4.2\, \mathrm{K}$ is shown as a function of $V_{\textrm{g}}$ between $G=3.5(2e^2 /h)\; \textrm{and}\; 6.3(2e^ 2/h)$ to show the conductance plateaus.
The inset shows the overall behavior of the conductance as a function of $V_{\textrm{g}} $.}
\label{figure:001}
\end{figure*}

Figure \ref{figure:001}(a) shows the heterostructure consisting of the following layers on an InP substrate: $\textrm{i-In}_{0.52}\textrm{Al}_{0.48}\textrm{As}$ $(200 \, \mathrm{nm})$, $\textrm{n-In}_{0.52}\textrm{Al}_{0.48}\textrm{As}$ $(15 \, \mathrm{nm})$, $\textrm{i-In}_{0.52}\textrm{Al}_{0.48}\textrm{As}$ $(15 \, \mathrm{nm})$, $\textrm{i-InGaAsP}$ $(5 \, \mathrm{nm})$, $\textrm{i-In}_{0.8}\textrm{Ga}_{0.2}\textrm{As}$ $(10 \, \mathrm{nm})$, $\textrm{InGaAlAs}$ $(3 \, \mathrm{nm})$, and $\textrm{i-In}_{0.52}\textrm{Al}_{0.48}\textrm{As}$ $(25 \, \mathrm{nm})$.
2DEG formed in $\textrm{i-In}_{0.8}\textrm{Ga}_{0.2}\textrm{As}$ layer has the carrier density of $1.10\times 10^{12} \, \mathrm{cm}^{-2}$ with the electron mobility of $11.65 \, \mathrm{m}^2 /\mathrm{Vs} $ at $0.3 \, \mathrm{K}$.
Our QPC structure was fabricated from this heterostructure defined by in-plane side gates patterned using the electron beam lithography and the reactive ion etching.
The conductance and noise measurements were performed on the QPC with a gate voltage $V_{\textrm{g}}$ applied at liquid helium temperature (Fig.\ref{figure:001}(b)).
Figure \ref{figure:001}(c) shows the conductance $G$ of the QPC as a function of $V_{\textrm{g}}$.
While the relevance of the spin-dependent transport was observed in this system for the low conductance region below $\sim 2(2e^2/h)$ \cite{KohdaUnpub}, in this paper, we focus ourselves on the region between $V_{\textrm{g}} =-1.2\, \mathrm{V}$ and $-2\, \mathrm{V}$, where the conductance ranges between $3.5(2e^2 /h)$ and $6.3(2e^2 /h)$.

\begin{figure}[b]
\centering
\includegraphics{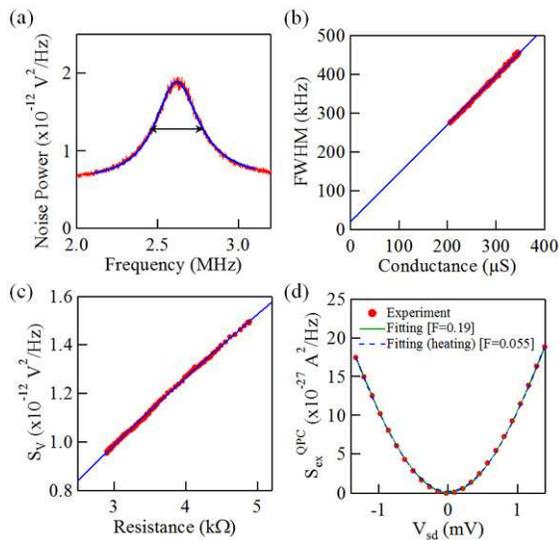}
\caption[left]{(a) Typical voltage noise peak around the resonance frequency.
The arrow shows the width (FWHM) of the peak.
(b) The width of the resonance peak as a function of the device conductance.
The curve is the result of the fitting as $\textrm{FWHM}=(1/2\pi C)(G+1/Z)$.
(c) $S_V $ as a function of the device resistance.
The curve is the result of the fitting as $S_V =A [(S_I +S_I^{{\textrm{ext}}})(1/R+1/Z)^2 +S_V^{{\textrm{ext}}}]$.
(d) Typical excess noise $S_{{\textrm{ex}}}^{{\textrm{QPC}}}$ as a function of $V_{{\textrm{sd}}}$ obtained at $V_{\textrm{g}} =-1.71 \mathrm{V}$.
The Fano factor is extracted from the curve fit by Eq.(\ref{eqn:001}) (solid curve).
The dashed curve is the simulated result with the heating effect.}
\label{figure:002}
\end{figure}

The shot noise, namely, the current fluctuation around its average $(I)$, was measured as follows \cite{NakamuraPRL2010,YamauchiPRL2011}.
The voltage noise across the sample on the resonant $(LC)$ circuit was referred to as an output signal of the home-made cryogenic amplifier followed by the room-temperature amplifier.
The resonance frequency was set to $2.6 \, \mathrm{MHz}$ with a bandwidth of $\sim 450 \, \mathrm{kHz}$.
The resultant time-domain signals were sampled by a digitizer and converted to spectral density via fast-Fourier transform.
A typical voltage noise spectral density around the resonance frequency was shown in Fig.\ref{figure:002}(a), from which the voltage noise power spectral density $S_V $ is deduced \cite{DiCarloRSI2006}.

In order to derive the current noise spectral density $S_I $ at the QPC, the noise measurement system was carefully calibrated.
Figure \ref{figure:002}(b) shows the full width at half maximum (FWHM) of the resonance peak as a function of the device conductance.
The impedance $Z$ originating from the $LC$ circuit and the capacitance $C$ at the resonance frequency were extracted from linear fitting as $\textrm{FWHM}=(1/2\pi C)(G + 1/Z)$, respectively \cite{DiCarloRSI2006,HashisakaPRB2008}, to yield $Z=63 \, \mathrm{k\Omega }$ and $C=130 \, \mathrm{pF}$.
In addition, when the cryogenic amplifier involves finite voltage and current noises $(S_{V}^{\textrm{ext}}\; \textrm{and}\; S_{I}^{\textrm{ext}})$, the observed $S_V $ for the sample with the circuit resistance $R$ is related to $S_I $ as $S_{V}=A[(S_I +S_{I}^{\textrm{ext}}) (1/R+1/Z)^2 +S_{V}^{\textrm{ext}}]$.
Here, $A$ is the total gain of the amplifiers.
Resistance $R$ consists of that of the QPC, the Hall bar, and the contacts.
By changing the QPC resistance by tuning $V_{\textrm{g}}$, we obtain $S_{V} $ as a function of $R$ as shown in Fig.\ref{figure:002}(c), from which $A=1.3\times 10^6 \, \mathrm{V^2 /V^2 }$, $S_{V}^{\textrm{ext}}=5\times 10^{-20} \, \mathrm{V^2 /Hz}$, and $S_{I}^{\textrm{ext}}=4\times 10^{-27} \, \mathrm{A^2 /Hz}$ are deduced.
Finally, to derive the shot noise and the QPC resistance from the observed $S_I$ and $R$, the contribution of the Hall bar and the contact resistances was taken into account.

\begin{figure}
\centering
\includegraphics{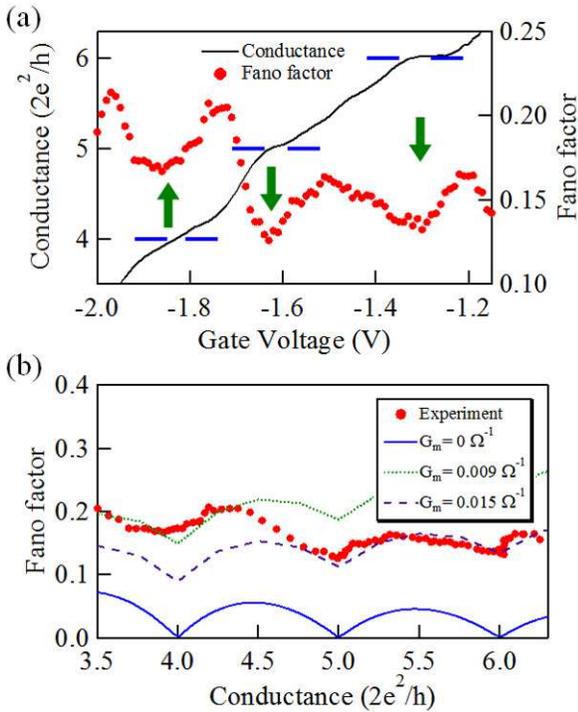}
\caption{(a) Fano factor as a function of $V_{\textrm{g}}$ (dot) and QPC conductance (curve).
(b) The Fano factor plot vs. the QPC conductance (dot).
The solid curve shows the theoretical Fano factor (see Refs. \cite{BlanterPR2000,ButtikerPRB1992,MartinPRB1992}).
Two dashed curves are the results of the simulation with the Fano factor including the heating effect $(G_{\textrm{m}} =0.009\; \textrm{and}\; 0.015\, \mathrm{\Omega^{-1}})$.}
\label{figure:003}
\end{figure}

Figure \ref{figure:002}(d) shows the typical excess noise $S_{ex}^{\textrm{QPC}}$ obtained at $V_{\textrm{g}} =-1.71 \, \mathrm{V}$ as a function of $V_{{\textrm{sd}}}$, where $G$ corresponds to $4.5(2e^ 2/h)$.
$S_{{\textrm{ex}}}^{\textrm{QPC}}$ is the current noise at the QPC whose thermal noise is subtracted.
$S_{\textrm{ex}}^{\textrm{QPC}}$ shows a parabolic behavior around $|V_{{\textrm{sd}}}|<2k_{\textrm{B}} T/e\sim 0.8 \, \mathrm{mV}$ for $T\sim 4.2 \, \mathrm{K}$, which is the crossover from thermal to shot noise ($k_{\textrm{B}} $ is the Boltzmann constant).
At higher $|V_{{\textrm{sd}}}|$, the excess noise is linearly dependent on $V_{{\textrm{sd}}}$.
We performed the numerical fitting to obtain the Fano factor $F$ by using the following conventional
formula \cite{BlanterPR2000}:
\begin{equation}
S_I =4k_{\textrm{B}} T(1-F)\frac{dI}{dV_{{\textrm{sd}}}}+2eIF\mathrm{coth}\left( \frac{eV_{{\textrm{sd}}}}{2k_{\textrm{B}} T} \right)
\label{eqn:001}
\end{equation}
The solid curve in Fig.\ref{figure:002}(d) is the result of the fitting, which yields $F=0.19$.
This Fano factor is deviated from that expected theoretically ($F=0.055$), which we discuss later [see dashed curve in Fig. 2(d)].
However, for the moment, we first look at the qualitative aspect of what the experimental result tells.
Figure \ref{figure:003}(a) shows the Fano factor as a function of $V_{\textrm{g}}$, which reduces at the conductance plateaus of $N(2e^2 /h)\; (N=4, 5, \textrm{and} \, 6)$.
Importantly, although the plateaus are not as perfectly flat as those usually observed in the InGaAs-based QPCs, the reduction of $F$ at each plateau is unambiguous.
This is the evidence that the shot noise suppresses due to the quantized coherent channel formation in the InGaAs-based QPC.

Ideally, no electrons can be reflected at the QPC at the conductance plateaus, leading to the absence of the shot noise, namely, $F=0$.
Our observation, however, shows that the Fano factor remains finite even at the plateaus.
Figure \ref{figure:003}(b) represents the Fano factor as a function of $G$.
The solid curve is the theoretical Fano factor, which is predicted to be $F=\Sigma_n T_n (1-T_n )/\Sigma_n T_n $, where $T_n $ indicates the transmission probability of the $n$-th quantized channel \cite{BlanterPR2000,ButtikerPRB1992,MartinPRB1992}.
Clearly, the experimentally obtained $F$ is larger than the theoretical prediction.

There are three possibilities to be responsible for the observed enhancement, namely, the electron heating, the channel mixing, and the $1/f$ noise.
Among them, the first one is the most likely.
We follow the model\cite{KumarPRL1996} that treats the electron heating as the diffusion by Wiedemann-Franz thermal conduction of the heat flux on both sides of the QPC and of Joule heating in the reservoirs, with the assumption that the ohmic contacts thermalized to the lattice at the system temperature $T$.
In this model, the electron temperature ($T_{\textrm{e}} $) is expressed as $T_{\textrm{e}} =T\sqrt{1+(24/\pi^2 )(G/G_{\textrm{m}})(1+2G/G_{\textrm{m}})(eV_{{\textrm{sd}}}/2k_{\textrm{B}}T)^2}$.
$G_{\textrm{m}}$ is the parallel conductance of the reservoirs that connect to the QPC.
Figure \ref{figure:002}(d) shows the simulated shot noise with $F=0.055$ and $G_{\textrm{m}}=0.0083\, \mathrm{\Omega^{-1}}$, which almost overlaps with the result without heating effect with $F=0.19$.
The dashed curves in Fig.\ref{figure:003}(b) are the simulation results with $G_{\textrm{m}}=0.009\; \textrm{and}\; 0.015\, \mathrm{\Omega^{-1}}$, which typically gives $T_{\textrm{e}}-T\sim 1\, \mathrm{K}$ at $V_{{\textrm{sd}}}=1\, \mathrm{mV}$.
Thus, the observed Fano factor can be explained by choosing $G_{\textrm{m}}$ around these values, which are consistent with the previous report \cite{KumarPRL1996}.
Although a single fixed value of $G_{\textrm{m}}$ cannot explain the whole experimental result, the channel mixing would explain it but a quantitative treatment including the mixing effect is difficult \cite{KumarPRL1996}.
We also note that a satisfying agreement between the experimental result and the simulation in Fig.\ref{figure:002}(d) strongly suggests the irrelevance of the $1/f$ noise contribution; if the $1/f$ noise would be significant, the excess noise would increase in proportional to $\sim V_{{\textrm{sd}}}^2$ so that the simulation could not explain the experimental result.
Also, usually, the contribution of the $1/f$ noise would be very drastic as reported before \cite{DekkerPRL1991}.

To conclude, we found that the shot noise is reduced at the conductance plateau of the QPC on the InGaAs/InGaAsP heterostructure, which is the direct evidence of the coherent quantized channel formation in this system.
The deviation of the Fano factor from the theory was mainly attributed to the electron heating effect.
Our achievement suggests that what has been realized in GaAs-based materials is also applicable in InGaAs systems, which, unlike GaAs-based systems, uniquely possess large $g$-factor and strong SOI.
Since the shot noise is apowerful tool to study quantum spin transport, further experimental efforts, for example, testing various theoretical predictions \cite{FevePRB2002,EguesPRL2002,AvishaiPRL2010,GovernalePRB2003}, will open up the possibilities of InGaAs-based devices in quantum information and spintronics fields.

This work is partially supported by JSPS Funding Program for Next Generation World-Leading Researchers.

\end{document}